\documentclass[apj]{emulateapj}

\begin{document}

\title{The Impact of Hot Jupiters on the Spin-down of their Host Stars}

\author{O. Cohen\altaffilmark{1}, J.J. Drake\altaffilmark{1}, V.L. Kashyap\altaffilmark{1}, 
I.V. Sokolov\altaffilmark{2}, and T.I. Gombosi\altaffilmark{2}}

\altaffiltext{1}{Harvard-Smithsonian Center for Astrophysics, 60 Garden St. Cambridge, MA 02138}
\altaffiltext{2}{Center for Space Environment Modeling, University of Michigan, 2455 Hayward St., 
Ann Arbor, MI 48109}

\begin{abstract}

We present a numerical Magnetohydrodynamic (MHD) study of the dependence of stellar mass and angular momentum- loss 
rates on the orbital distance to close-in giant planets.
We find that the mass loss rate drops by a factor of $\approx$1.5-2, while the angular momentum 
loss rate drops by a factor of $\approx$4 as the distance decreases past the Alfv\'en surface. This reduction in 
angular momentum loss is due to the interaction between the stellar and planetary Alfv\'en surfaces, which modifies the 
global structure of the stellar corona and stellar wind on the hemisphere facing the planet, as well as the opposite 
hemisphere. The simulation also shows that
the magnitude of change in angular momentum loss rate depends mostly on the strength of the planetary magnetic field 
and not on its polarity. The interaction however, begins at greater separation if the overall field topology of the 
star and the planet are of anti-aligned. Our results are consistent with evidence for excess 
angular momentum in stars harboring close-in giant planets, and show that the reduction in wind-driven angular momentum 
loss can compete with, and perhaps dominate, spin-up due to tidal interaction.

\end{abstract}



\section{INTRODUCTION}
\label{sec:Intro}

Nearly two decades after the first discovery of a planet orbiting a main-sequence star outside of the 
solar system \citep{mayor95}, hundreds of other exoplanets have been observed \citep{exoplanets03}. 
Many of these planets have masses in the Jupiter class, and orbit their host star at distances of ${\lesssim}10$ stellar radii.
Such massive close-in planets are also called ``hot Jupiters''.

Planets that are far out in the stellar system cannot significantly affect their 
host star via tidal (gravitational) interaction. They also cannot affect the star via magnetic 
interaction (even if the planet has a strong internal magnetic field)
since the super-Alfv\'enic stellar wind prevents the propagation of information
back from the planet to the star along the field lines; the star is thus essentially
blind to the existence of the planetary magnetic field. The conceptual boundary which separates 
the sub- and super-Alfv\'enic regimes is called the ``Alfv\'en surface'' located at a radial distance from the star 
known as the "Alfv\'en radius".  This surface is defined 
as the collection of critical points at which the wind flow speed, $u$, is equal to the Alfv\'en speed, 
$u_A$, or in other words, the Alfv\'enic Mach number, $M_A=u/u_A=1$.  
In the case of close-in planetary orbits at distances comparable to or less than the stellar Alfv\'en radius, 
magnetic interactions are more easily facilitated and can, at least in principle, induce a feedback on the star 
via star-planet magnetic interaction.

Signatures of star-planet interaction have been observed through local modulations in the Ca II K emission 
line---an indicator of chromospheric activity---in several planetary systems 
\citep[][see also \citet{Hartman10}]{shkolnik05a,shkolnik05b,shkolnik08}. In addition, an apparent enhancement 
in stellar activity in stars with close-in planets compared to stars with planets located at greater orbital 
radii appears in the form of an overall increase in observed X-ray \citep[][see, however, 
\citealt{Poppenhaeger10}]{kashyap08} and UV \citep{shkolnik10} flux.  

Recently, \cite{Pont09} performed an empirical study between stellar rotation rate, orbital semimajor axis, 
stellar and planetary radii, and the planet-to-star mass ratio. This study suggests that stars that harbor 
hot Jupiters rotate faster than stars with planets 
located at larger orbital separation, or stars without planets at all. \cite{Pont09} followed \citep{MardlingLin02} 
in suggesting that the excess in stellar angular momentum is due to tidal torques that synchronize the stellar rotation 
with the planetary orbital motion.  It is also possible to view this result as a reduction in the angular momentum lost 
from stars with hot Jupiters, rather than as the host stars being spun up. \citep{Lanza10} suggests that in many observed 
planetary systems the synchronization time-scale is too long for tidal interaction to cause this excess in stellar 
angular momentum. Instead, he proposes that the angular momentum excess is due to star-planet magnetic interaction 
that leads to a reduction of magnetic breaking and angular momentum loss to the stellar wind \citep{weberdavis67}. 
He concludes that both mechanisms can take part in the process of angular momentum transfer and that the dominant of 
the two mechanisms depends on the particular set of stellar-planetary parameters, as well as 
the stellar type. 

In this report, we present MHD numerical simulations of the star-planet 
interaction, where we focus on calculating the stellar angular momentum loss rate for a representative 
set of stellar and planetary parameters as a function of the planetary orbital radius.  We describe the simulation details in \S\ref{sec:Simulation},
describe the results and in \S\ref{sec:Results}.  We summarize in \S\ref{sec:Conclusions}.


\section{SIMULATION}
\label{sec:Simulation}

The simulations described  here were performed in the same manner as in \citep{Cohen09b}.  The model used 
solves the set of MHD equations, based on the surface distribution of the stellar magnetic field, 
and provides a self-consistent, steady-state solution for the stellar corona and stellar wind 
\citep{Cohen07}. The planet is represented by a dipole magnetic field with boundary 
conditions typical of Jupiter. Since the simulation provides the solution for the density, 
velocity, magnetic field, and pressure, we can calculate the realistic Alfv\'en surface and 
hence the stellar angular momentum loss rate (see \citealt{cohen09a} for a detailed description 
of this procedure). 

We adopt some of the stellar properties of the well observed HD~189733 system
($R_\star=0.76R_\odot$, $M_\star=0.82M_\odot$ \citep{exoplanets03}, and
$P_{rot}=11.95d$ \citep{Fares10}). \citep{Fares10} have recently investigated the surface magnetic field distribution 
of HD 189733; here we adopt an idealized dipolar stellar magnetic field in order not to complicate any modulation of 
model results with planetary distance. The planetary field is assumed to be dipolar as well, with 
equatorial field of $B_p=1\;G$. The choice of a field somewhat lower than that of  Jupiter is based on the expected slower 
rotation of close-in planets  as a result of tidal locking \citep{sanches-lavega04}, and the observed inverse correlation 
between planetary magnetic field strength and rotation period \citep[e.g.][]{durand-Manterola09}.

In addition to the stellar Alfv\'en surface, we expect the planet to possess an analogous  surface due to planetary outflow. 
In this simulation, we mimic the planetary outflow by introducing the planet as a mass source for the plasma 
outflow, where the flow rate depends on the particular choice of the planetary boundary conditions. 
The planetary radius used here is $R_p=1.95~R_{J}$ ($R_J$ is Jupiter's radius) with boundary conditions for the density, 
$n_{p}=10^7\;cm^{-3}$, and temperature, $T_{p}=10^4\;K$. This particular choice of parameters produces a planetary 
outflow of about $50\;km\;s^{-1}$. 

Here, we study modulations in stellar angular momentum loss rate as a function of the semimajor axis, $a$. 
Once the stellar Alfv\'en surface is defined, we calculate the mass and angular momentum loss rates which 
depend on the topology of this surface. We explore values of $a$ ranging from $12R_\star$ to $3R_\star$, and calculate the loss 
rates for each case.  We omit the case of $a=5R_\star$, since at this distance the planet 
is located exactly at the stellar Alfv\'en surface. This causes the stellar Alfv\'en surface to be an 
open surface with the grid space of the simulations and the integration required for the loss rate calculations is ill-conditioned.
We also test how the orientation of the planetary magnetic field affects 
the loss rates by defining a planetary dipole magnetic field which is aligned or anti-aligned 
to the stellar dipolar magnetic field.


\section{RESULTS \& DISCUSSION}
\label{sec:Results}

\begin{figure*}[h!]
\centering
\includegraphics[width=7.in]{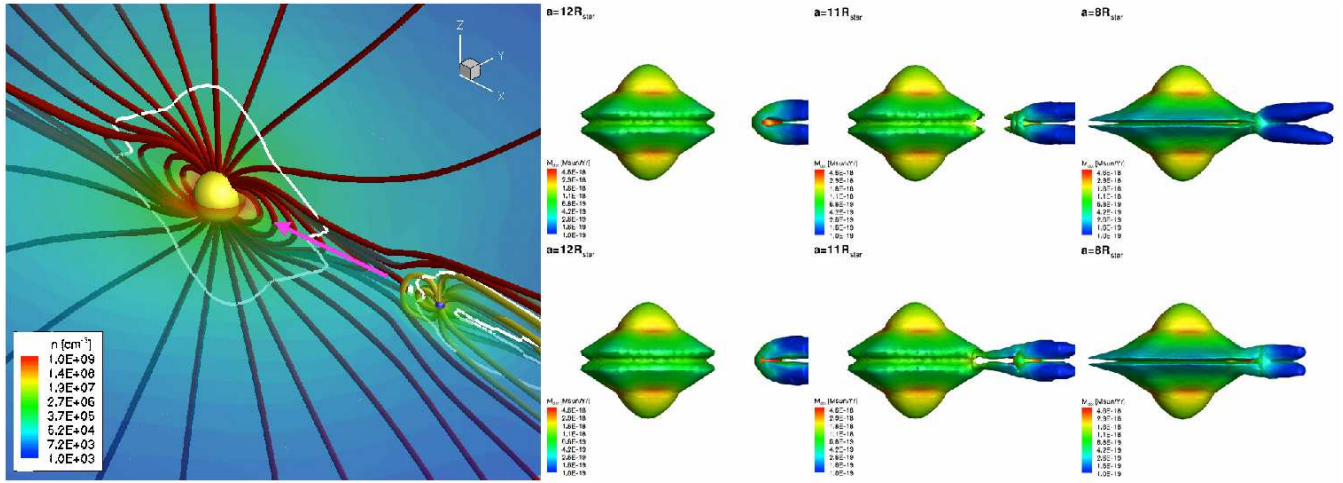}
\caption{Left: Steady-state solution for the stellar corona and the planet for the $a=12R_\star$, aligned dipole 
case. The star and the planet are shown as yellow and blue spherical shades, respectively. Red lines 
represent the stellar magnetic field lines, while yellow lines represent the planetary magnetic field lines. 
Background color contours are of number density and the Alfv\'en surfaces of the two bodies are shown 
as solid white lines. The pink arrow marks the direction at which the planetary 
location has been modified with distance. Right: Stellar and planetary Alfv\'en surfaces for the cases of 
$a=12,11,8R_\star$ (left to right, respectively) colored with contours of the local mass flux at each point. 
Top row shows the solution for the aligned planetary dipole case while the bottom row shows the solution for the 
anti-aligned planetary dipole case.}
\label{fig:f1}
\end{figure*}

The MHD solution for the particular case of an aligned planetary dipole and semimajor axis $a=12R_\star$ is illustrated in the 
left panel of Figure~\ref{fig:f1}.  Also shown 
with an arrow is the range of semimajor axis investigated.
The six panels on the right of Figure~\ref{fig:f1} show the stellar and planetary 
Alfv\'en surfaces for different values of $a$ and for aligned and anti-aligned dipoles.   

\begin{figure*}[h!]
\centering
\includegraphics[width=7.in]{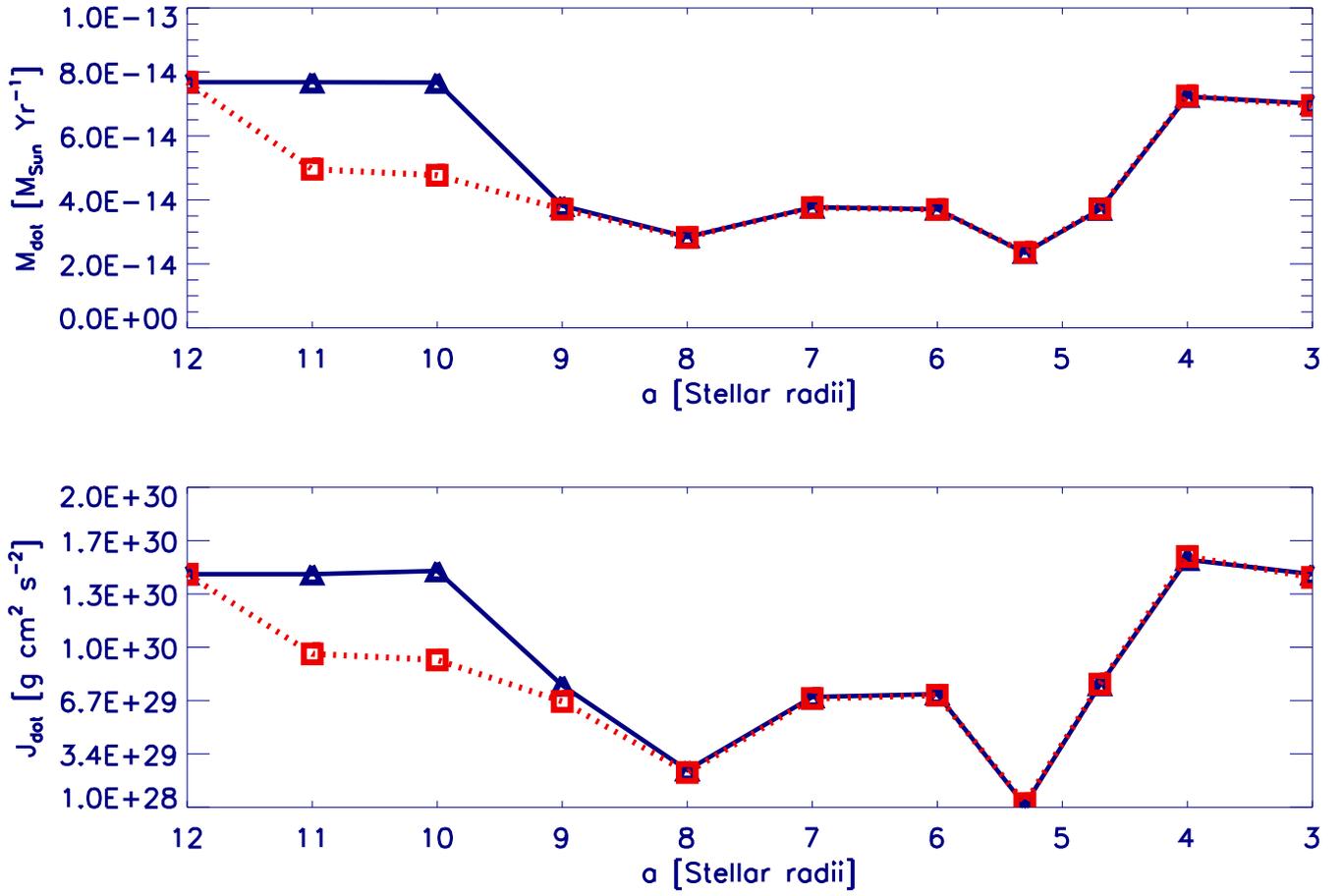}
\caption{Mass loss rates, $M_{dot}$, (top, units of solar mass per year) and angular momentum loss rates, 
$J_{dot}$, (bottom, units of $[g\;cm^2\;s^{-2}]$) for the case of aligned planetary dipole (solid blue line, triangles) and 
anti-aligned planetary dipole (dashed red line, squares).}
\label{fig:f2}
\end{figure*}

\begin{figure*}[h!]
\centering
\includegraphics[width=7.in]{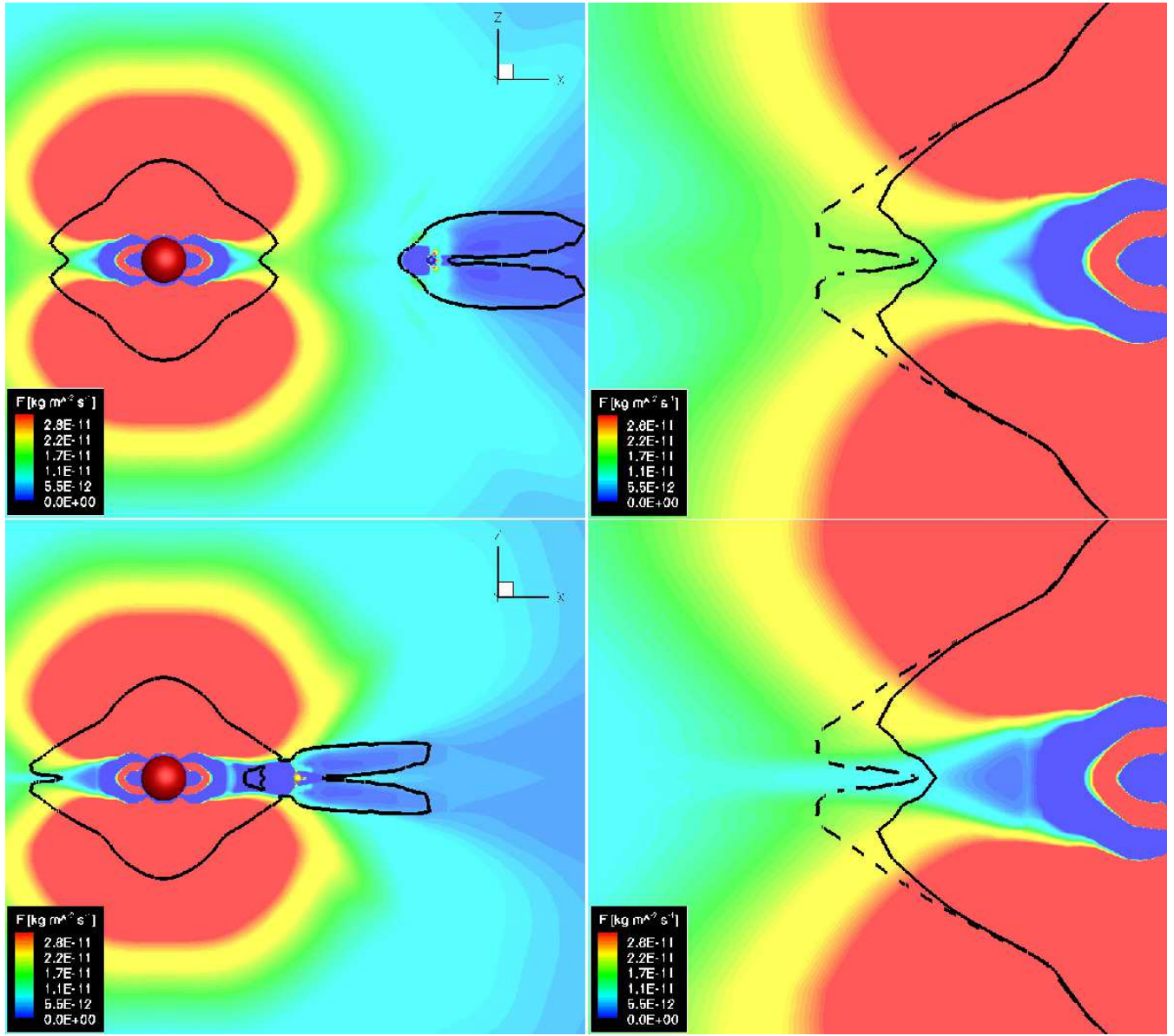}
\caption{Mass flux distribution (color contours) and the location of the Alfv\'en surface (black line) displayed on the 
y=0 (star-planet) meridional plain. 
The solution for the $a=12R_\star$ case is shown on the top panels while the solution for the $a=6R_\star$ case is shown on the bottom panels. 
Left panels show the global solution with the star as a red sphere, and the planet as a blue sphere ($a=12R_\star$) or yellow sphere 
($a=6R_\star$). The right panels show a zoom on the vicinity of the stellar corona hemisphere which is 
not occupied by the planet. In both right panels, the Alfv\'en surface for the $a=12R_\star$ case is shown as a solid line, while the 
Alfv\'en surface for the $a=6R_\star$ case is shown as a dashed line.}
\label{fig:f3}
\end{figure*}

The results for the mass and angular momentum loss rates as a function of $a$ are shown in Figure~\ref{fig:f2}. 
It can be seen that the mass loss rate drops by about a factor of 1.5-2 at $a=9R_\star$ for the aligned 
case and at $a=11R_\star$ for the anti-aligned case. The trends in the plot for the angular momentum loss 
rate are similar, but here the change reaches a factor of 4 at $a=8R_\star$. The reason for this change is 
clearly seen in Figure~\ref{fig:f3}, where we show the mass flux distribution, as well as the location of the 
Alfv\'en surface over the star-planet meridional plain for the cases of $a=12R_\star$ and $a=6R_\star$. In the case 
of $a=12R_\star$, the front of the planetary Alfv\'en surface is beyond the stellar one, so that the existence 
of the planet does not affect the structure of the stellar corona. In the case of $a=6R_\star$, the planet is 
located inside the stellar Alfv\'en surface, so the stellar corona and stellar wind the ``feel'' the obstacle 
(i.e. the planet and its magnetosphere), and the large-scale structure of the corona is disrupted. In particular, 
both on the planetary side and on the opposite side, the mass flux near 
the equatorial region is significantly reduced due to the existence of the planet. This explains the 50\% reduction 
in mass flux when the planet affects the corona. The fact that the coronal side opposite to the planet is modified 
demonstrates that the effect of the planet on the corona is global. We would like to point out that another reference simulation 
with different set of stellar and planetary parameters resulted in the same trend, so that this effect does not depend 
on the particular choice of parameter set. In addition, an initial result 
from a simulation of the HD 189733 planetary system, which includes realistic stellar magnetic field, as well as 
a planetary orbital motion resulted in a total mass flux of $6\cdot 10^{-14}\;M_\odot\;yr^{-1}$, which is close to the 
value of $4\cdot 10^{-14}\;M_\odot\;yr^{-1}$ obtained from this simulation for $a=9R_\star$ (as observed for HD189733).

Since in our case the star and the planet are tidally locked, the planet blocks a significant 
amount of the stellar wind, and prevents the stellar wind from opening up field lines and escaping \citep{Cohen09b}. 
For the particular set of parameters used in our simulations, the planetary magnetosphere has a large enough 
cross-section to block the stellar wind over a significant area. A planet without a significant internal magnetic 
field would not only be too small to affect the mass flux through the Alfv\'en surface, but would probably suffer 
from a strong erosion by the stellar wind. In our simulations, the radius of the cross section of the planetary 
magnetosphere is not larger than about 1-2 stellar radii. Therefore, the area which is blocked by the planet is 
not more than about 25-30\% of the total area of the stellar Alfv\'en surface. Nevertheless, the whole coronal 
hemisphere that faces the planet is affected, so that the stellar mass loss rate decreases by approximately a 
factor of 1.5-2 as a result of the interaction with the planetary magnetosphere. 

The four right panels of Figure~\ref{fig:f2} show that the planet begins to affect the corona when the planetary 
Alfv\'en surface, (and {\it not the planet itself}) starts to interact with the stellar one. They also shows that 
the two surfaces merge at greater separation for the case of an anti-aligned planetary dipole as compared to the 
aligned case. This explains the smaller value of $a$ at which a drop in mass and angular momentum loss rates compared  to 
the aligned dipole case is seen. This merging of the surfaces at smaller $a$ is due to particular magnetic field lines that 
connect the planetary magnetic field and the stellar one. 

The amount of  change in the loss rates depends mostly on the size of the planetary magnetosphere, 
which is determined by the particular choice of planetary parameters. The alignment of the planetary field has only 
a minor effect on the change in angular momentum loss. 
This is due to the fact that the wind is blocked primarily 
by the magnetic pressure of the planetary field, which depends on the magnitude of the field but not on its 
polarity.  It is most likely that a tilted planetary dipole will produce only slightly different result, depending on the tilt 
angle. 

\citet{Pont09} presented empirical evidence for ``excess'' rotation in stars hosting close-in planets. 
This was interpreted in terms of tidal forces acting towards spin-orbit synchronization and transfer of orbital to 
rotational angular momentum.  While such angular momentum exchange doubtless takes place, the MHD models presented 
here demonstrate that the stellar angular momentum loss rate through magnetized winds also likely plays a significant role. 
Close-in planets reduce this angular momentum loss and reduce stellar spin-down.  The effect is large enough---a factor 
of up to 4---that it could dominate the observed rotation excesses.  As noted by \citet{Pont09}, spin-up by tidal effects
 requires substantial orbital decay since the dissipation of the protoplanetary disk.  
The reduced magnetic braking effect found here would imply that much less severe orbital decay would be required.

Full understanding of the distribution of planetary orbits and stellar rotation rates likely requires consideration of 
both tidal and magnetic angular momentum loss and exchange.  Improvements in the MHD modeling presented can be made at 
the expense of substantial computational cost by including time-dependent effects of the star-planet interaction using 
more realistic surface magnetic field distributions.


\section{Conclusions}
\label{sec:Conclusions}

In this report we calculate variations in stellar angular momentum loss rate of a star hosting 
a hot Jupiter planet as a function of the planetary orbital distance. We show that once the stellar 
and planetary Alfv\'en surfaces interact with each other, the stellar wind topology in the hemisphere 
facing the planet changes and the angular momentum loss to the wind decreases. The simulation also 
shows that this interaction begins at a greater distance if the overall field topology of the star 
and the planet are opposite, though the magnitude of change in angular momentum loss rate depends mostly on the 
size of the planetary magnetosphere.   This reduction in angular momentum loss for close-in planets likely contributes  
to, and perhaps even dominates, the observed excess rotation found in stars with close-in planets. 


\acknowledgments
We thank an unknown referee for his/hers useful comments. OC is supported by SHINE through NSF ATM-0823592 grant, and by NASA-LWSTRT Grant NNG05GM44G.
JJD and VLK were funded by NASA contract NAS8-39073 to the {\it Chandra X-ray Center}.
Simulation results were obtained using the Space Weather Modeling
Framework, developed by the Center for Space Environment Modeling, at the University of Michigan with funding
support from NASA ESS, NASA ESTO-CT, NSF KDI, and DoD MURI.



\begin{thebibliography}{19}
\expandafter\ifx\csname natexlab\endcsname\relax\def\natexlab#1{#1}\fi

\bibitem[{{Cohen} {et~al.}(2009{\natexlab{a}}){Cohen}, {Drake}, {Kashyap}, \&
  {Gombosi}}]{cohen09a}
{Cohen}, O., {Drake}, J.~J., {Kashyap}, V.~L., \& {Gombosi}, T.~I.
  2009{\natexlab{a}}, Astrophys. J., 699, 1501

\bibitem[{{Cohen} {et~al.}(2009{\natexlab{b}}){Cohen}, {Drake}, {Kashyap},
  {Saar}, {Sokolov}, {Manchester}, {Hansen}, \& {Gombosi}}]{Cohen09b}
{Cohen}, O., {Drake}, J.~J., {Kashyap}, V.~L., {Saar}, S.~H., {Sokolov}, I.~V.,
  {Manchester}, W.~B., {Hansen}, K.~C., \& {Gombosi}, T.~I. 2009{\natexlab{b}},
  Astrophys. J., 704, L85

\bibitem[{{Cohen} {et~al.}(2007){Cohen}, {Sokolov}, {Roussev}, {Arge},
  {Manchester}, {Gombosi}, {Frazin}, {Park}, {Butala}, {Kamalabadi}, \&
  {Velli}}]{Cohen07}
{Cohen}, O., {Sokolov}, I.~V., {Roussev}, I.~I., {Arge}, C.~N., {Manchester},
  W.~B., {Gombosi}, T.~I., {Frazin}, R.~A., {Park}, H., {Butala}, M.~D.,
  {Kamalabadi}, F., \& {Velli}, M. 2007, \apjl, 654, L163

\bibitem[{{Durand-Manterola}(2009)}]{durand-Manterola09}
{Durand-Manterola}, H.~J. 2009, \planss, 57, 1405

\bibitem[{{Fares} {et~al.}(2010){Fares}, {Donati}, {Moutou}, {Jardine},
  {Grie{\ss}meier}, {Zarka}, {Shkolnik}, {Bohlender}, {Catala}, \&
  {Cameron}}]{Fares10}
{Fares}, R., {Donati}, J., {Moutou}, C., {Jardine}, M.~M., {Grie{\ss}meier},
  J., {Zarka}, P., {Shkolnik}, E.~L., {Bohlender}, D., {Catala}, C., \&
  {Cameron}, A.~C. 2010, \mnras, 735

\bibitem[{{Hartman}(2010)}]{Hartman10}
{Hartman}, J.~D. 2010, \apjl, 717, L138

\bibitem[{{Kashyap} {et~al.}(2008){Kashyap}, {Drake}, \& {Saar}}]{kashyap08}
{Kashyap}, V.~L., {Drake}, J.~J., \& {Saar}, S.~H. 2008, Astrophys. J., 687,
  1339

\bibitem[{{Lanza}(2010)}]{Lanza10}
{Lanza}, A.~F. 2010, Astronomy \& Astrophysics, 512, A77

\bibitem[{{Mardling} \& {Lin}(2002)}]{MardlingLin02}
{Mardling}, R.~A., \& {Lin}, D.~N.~C. 2002, \apj, 573, 829

\bibitem[{{Mayor} {et~al.}(2003){Mayor}, {Naef}, {Pepe}, {Queloz}, {Santos}, \&
  {Udry}}]{exoplanets03}
{Mayor}, M., {Naef}, D., {Pepe}, F., {Queloz}, D., {Santos}, N., \& {Udry}, S.
  2003, {The Geneva extrasolar planet search programmes},
  {http://exoplanets.eu}

\bibitem[{{Mayor} \& {Queloz}(1995)}]{mayor95}
{Mayor}, M., \& {Queloz}, D. 1995, Nature, 378, 355

\bibitem[{{Pont}(2009)}]{Pont09}
{Pont}, F. 2009, MNRAS, 396, 1789

\bibitem[{{Poppenhaeger} {et~al.}(2010){Poppenhaeger}, {Robrade}, \&
  {Schmitt}}]{Poppenhaeger10}
{Poppenhaeger}, K., {Robrade}, J., \& {Schmitt}, J.~H.~M.~M. 2010, \aap, 515,
  A98+

\bibitem[{{S{\'a}nchez-Lavega}(2004)}]{sanches-lavega04}
{S{\'a}nchez-Lavega}, A. 2004, Astrophys. J., 609, L87

\bibitem[{{Shkolnik}(2010)}]{shkolnik10}
{Shkolnik}, E. 2010, in Bulletin of the American Astronomical Society, Vol.~41,
  Bulletin of the American Astronomical Society, 531

\bibitem[{{Shkolnik} {et~al.}(2008){Shkolnik}, {Bohlender}, {Walker}, \&
  {Collier Cameron}}]{shkolnik08}
{Shkolnik}, E., {Bohlender}, D.~A., {Walker}, G.~A.~H., \& {Collier Cameron},
  A. 2008, Astrophys. J., 676, 628

\bibitem[{{Shkolnik} {et~al.}(2005{\natexlab{a}}){Shkolnik}, {Walker},
  {Bohlender}, {Gu}, \& {Kurster}}]{shkolnik05a}
{Shkolnik}, E., {Walker}, G.~A.~H., {Bohlender}, D.~A., {Gu}, P.-G., \&
  {Kurster}, M. 2005{\natexlab{a}}, Astrophys. J., 622, 1075

\bibitem[{{Shkolnik} {et~al.}(2005{\natexlab{b}}){Shkolnik}, {Walker},
  {Rucinski}, {Bohlender}, \& {Davidge}}]{shkolnik05b}
{Shkolnik}, E., {Walker}, G.~A.~H., {Rucinski}, S.~M., {Bohlender}, D.~A., \&
  {Davidge}, T.~J. 2005{\natexlab{b}}, Astronom. J., 130, 799

\bibitem[{{Weber} \& {Davis}(1967)}]{weberdavis67}
{Weber}, E.~J., \& {Davis}, L.~J. 1967, \apj, 148, 217

\end{thebibliography}


\clearpage

\end{document}